\documentclass[aps,prl,twocolumn,showpacs]{revtex4}
\usepackage{graphicx}
\usepackage{times}
\usepackage{latexsym}
\def\beq{\begin{equation}}
\def\eeq{\end{equation}}
\def\bey{\begin{eqnarray}}
\def\eey{\end{eqnarray}}

\def\kpc{\, {\rm kpc} }

\def\lsim{\mathrel{\raise.3ex\hbox{$<$\kern-.75em\lower1ex\hbox{$\sim$}}}}
\def\gsim{\mathrel{\raise.3ex\hbox{$>$\kern-.75em\lower1ex\hbox{$\sim$}}}}

\begin{document}

\title{MeV Dark Matter and Small Scale Structure}  
\author{Dan Hooper$^1$, Manoj Kaplinghat$^2$, Louis E.~Strigari$^2$ and Kathryn M.~Zurek$^3$}
\address{$^1$Theoretical Astrophysics, Fermi National Accelerator Laboratory, Batavia, IL  60510 \\ $^2$Center for Cosmology, Department of Physics and Astronomy, University of California, Irvine, CA  92697 \\ $^3$Phenomenology Institute, University of Wisconsin, Madison, WI  53706}

\date{\today}

\begin{abstract}

WIMPs with electroweak scale masses (neutralinos, etc.) remain in
kinetic equilibrium with other particle species until temperatures
approximately in the range of 10 MeV to 1 GeV, leading to the
formation of dark matter substructure with masses as small as
$10^{-4} \, M_{\odot}$ to $10^{-12} \, M_{\odot}$. However, if dark
matter consists of particles with MeV scale masses, as motivated by
the observation of 511 keV emission from the Galactic Bulge, such
particles are naturally expected to remain in kinetic equilibrium with the cosmic neutrino
background until considerably later times. This would lead to a strong
suppression of small scale structure with masses below about $10^7\, 
M_{\odot}$ to $10^4\, M_{\odot}$. This cutoff scale has important
implications for present and future searches for faint Local Group
satellite galaxies and for the missing satellites problem.

\end{abstract}
\pacs{95.35.+d;95.30.Cq; FERMILAB-PUB-07-064-A; MADPH-07-1482}
\maketitle

{\bf Introduction.} In the standard cosmology, featuring cold,
collisionless dark matter, structures form hierarchically with the
smallest mass objects forming first and progressively larger objects
forming via subsequent mergers and accretion. This paradigm has been
remarkably successful at describing the observed large scale
structure.  

Within this paradigm, the mass of the smallest dark matter halos 
depends on the mass of the dark matter particles, and on
the temperature at which they decouple kinetically from other particle
species. For a typical weakly interacting massive particle (WIMP) with
an electroweak scale mass, kinetic decoupling from the standard model
leptons occurs at a temperature in the range of roughly 10 MeV to 1
GeV, leading to the formation of structures with masses as small as
$10^{-4} \, M_{\odot}$ to $10^{-12}\, M_{\odot}$~\cite{wimp,loeb}. If
dark matter consists of particles which remain in kinetic equilibrium
with neutrinos until later times (lower temperatures), the smallest
dark matter halos will be considerably more massive than are predicted
for WIMPs with electroweak scale masses \cite{mangano}.   

Dark matter particles with MeV scale masses have been previously
motivated by the observation of 511 keV emission from the Galactic
Bulge~\cite{integral,511dark}. In particular, 
annihilating MeV dark matter can inject the required rate of positrons
into the Galactic Bulge, and also be produced in the early universe
with the measured dark matter abundance~\cite{511dark}. Dark matter in
the form of MeV mass scalars, $\phi$, annihilating through the
exchange of a light gauge boson, $U$, can accommodate these
requirements~\cite{511dark,scalar,fayet}. Constraints on this
scenario have been placed by colliders~\cite{drees,collider}, neutrino
experiments~\cite{boehmneutrino}, atomic physics
experiments~\cite{atomic}, observations of supernova 1987A~\cite{sn}, 
the 511 keV line width~\cite{beacom} and Big Bang Nucleosynthesis (BBN)~\cite{bbn}.

In this letter, we revisit the MeV dark matter scenario, and calculate
the resulting matter power spectrum. We find that if the $U$-boson's
couplings to neutrinos is similar to its couplings to electrons, the
matter power spectrum is suppressed on small scales, leading to an
absence of dark matter halos with masses below about $10^7 \,
M_{\odot}$ to $10^4\, M_{\odot}$. We show that this  suppression scale
is consistent with the region of parameter space where the MeV dark
matter has the correct relic abundance as determined by recent
cosmological observations \cite{wmap}.

{\bf The MeV Dark Matter Power Spectrum.}
To calculate the power spectrum for MeV dark matter, we start by
determining  the temperature at which kinetic decoupling occurs. The
squared amplitude for dark matter-neutrino elastic scattering is given
by  
\begin{equation}
|{\cal M_{\phi \nu}}|^2 = \frac{8 g^2_{U \phi \phi} g^2_{U \nu \nu}
 m^2_{\phi} E_\nu^2}{(t-m^2_U)^2} \bigg[1+\cos \theta \bigg], 
\end{equation}
where $t = -2 E_\nu^2 (1-\cos \theta)$, $m_\phi$ is the mass of the
dark matter particle, $m_U$ is the mass of the exchanged boson, and
$g_{U \phi \phi}$ and $g_{U \nu \nu}$ are that boson's couplings to
dark matter and neutrinos, respectively. This leads to an elastic scattering cross section (in the $m_U \gg E_{\nu}$ limit) of
\begin{equation}
\sigma_{\phi \nu} = \frac{g^2_{U \phi \phi} g^2_{U \nu \nu} E^2_{\nu}}{2 \pi m^4_U}.
\end{equation}

To determine the temperature of kinetic decoupling for the dark matter
particle, we solve the Boltzmann equation, including the $\phi-\nu$
collision term. The
resulting  equation~\cite{boltz} is 
\begin{equation}
\frac{d\!f({\vec p})}{d\!t}=\Gamma(T_\nu)(T_\nu m_\phi \nabla_{\vec
  p}^2+{\vec p}\cdot \nabla_{\vec p} + 3)f({\vec p})\,,
\label{eq:boltz}
\end{equation}
where $\Gamma=31\pi^3 g_{U\phi\phi}^2 g_{U\nu\nu}^2 T_\nu^6 /
(42m_U^4m_\phi)$ is the rate for the dark matter distribution function
to relax to its equilibrium value. An intuitive approximation to this
relaxation rate is $\dot{E}_k/E_k=(4\pi)^{-1} \int d\!E_\nu
d\!\Omega (d\!n_\nu/d\!E_\nu) (d\!\sigma_{\phi\nu}/d\!\Omega)
\delta\!E_k/E_k$ where $d\!n_\nu/d\!E_\nu$ is the differential number
density of all neutrinos, $E_k=|{\vec   p}|^2/2m_\phi$ is the kinetic
energy of dark matter particles and $\delta\!E_k$ is the kinetic
energy transferred per collision. This approximation yields 
$15.2g_{U\phi\phi}^2 g_{U\nu\nu}^2 T_\nu^6 / (m_U^4m_\phi)$ for
$E_k=3T_\nu/2$, which is a factor of about 1.5 smaller the exact
result.  

Eq.~\ref{eq:boltz} is solved by a Boltzmann distribution
with a temperature for the dark matter particle that scales as $T_\nu$
during the strongly coupled regime and as $T_\nu^2$ after
decoupling. We define the kinetic decoupling
temperature as $T_{\rm kd} = T_\nu$, such that $\Gamma(T_\nu)=H(T_\nu)
\approx 5.97\sqrt{G_N}T_\nu^2$. This gives us  
\begin{equation}
T_{\rm kd} = 2.1 \, {\rm keV} \,  \frac{m_U}{{\rm MeV}} \,
    \left(\frac{m_\phi}{{\rm MeV}}\right)^{1/4}
    \left(\frac{10^{-6}}{g_{U\phi\phi}g_{U\nu\nu}} \right)^{1/2}\,. 
\end{equation}
Using the formalism of Ref.~\cite{loeb}, we can use this result to
calculate the power spectrum  of MeV dark matter.

The interactions and the subsequent decoupling of the dark matter
particles leads to the damping of the matter power spectrum. This
results from three distinct processes. First, the coupling of the
dark matter to other particle species introduces damped oscillatory
features~\cite{loeb}. This is the dominant effect for the case of
WIMPs with electroweak scale masses. Second, after decoupling, the
free-streaming of the dark matter particles further suppresses the
power spectrum. For MeV dark matter, this effect dominates
for the viable region of parameter space where $T_{\rm kd} \gtrsim
{\rm keV}$. Third, as neutrinos kinetically decouple from the dark 
matter they begin to free-stream and damp the power spectrum
further. This effect, however, is subdominant.  

\begin{figure}
\resizebox{8.0cm}{!}{\includegraphics{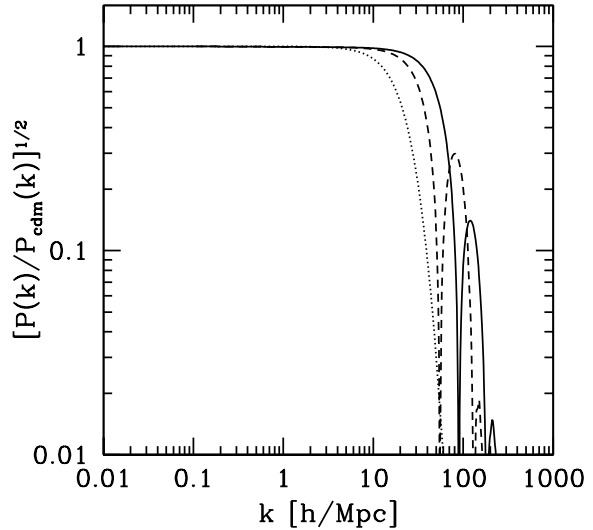}}
\caption{The effect of keV scale kinetic decoupling on the matter
  power spectrum, as predicted in MeV dark matter. Shown
  are results for a 1 MeV dark matter particle with a 10 keV (solid)
  and 1.0 keV (dashed) kinetic decoupling temperature. The dotted line
  denotes the limit relevant for warm dark matter, as inferred from
  observations of the lyman-alpha forest \cite{lymanalpha}.} 
\label{tf}
\end{figure}

In Fig.~\ref{tf}, we show the effect on the matter power spectrum
of MeV dark matter as compared to that for the
standard cold dark matter case. Large wavenumbers are strongly
suppressed, resulting in reduced number of small dark matter
halos. Also shown in the figure as a dotted curve is the
(strictest) limit found for the case of warm dark matter from
observations of the lyman-alpha forest 
\cite{lymanalpha}. 

For $T_{\rm kd} \gtrsim {\rm keV}$, the scale at which the power spectrum is truncated is closely related to the free-streaming scale, 
\begin{equation}
k_f^{-1} = 2.5 \, \kpc \, \left({{\rm keV} \over T_{\rm kd}}\right)^{1/2}
\left({{\rm MeV} \over m_\phi}\right)^{1/2}
\ln(4a_{\rm EQ}/a_{\rm kd})\,,
\end{equation}
where $a_{\rm{kd}}$ and $a_{\rm{EQ}}$ are the scale factors at
decoupling and matter-radiation equality, respectively. The
suppression of the dark matter power spectrum on scales smaller than
$k_f^{-1}$, in turn, leads to a cutoff in the mass function of dark
matter halos. Compared to the case with no cutoff, one would find a
paucity of halos with masses less than roughly $4\pi(\pi/k_f)^3 \,
\rho_M / 3$, where $\rho_M$ is the present cosmological matter density. To obtain a more accurate estimate, we find the mass at which
the expected number of dark matter halos falls by a factor of $e$
compared to the prediction for dark matter particles with electroweak scale masses. We calculate the
mass function of dark matter halos using the Press-Schechter
prescription. We note that the validity of this prescription for
power spectra with sharply truncated power (as found in our scenario) has
not been conclusively  demonstrated. Nevertheless, the cutoff mass
derived here is useful in the sense that it highlights the mass scale
below which we expect deviations from the predictions of standard cold dark matter. We find this cutoff mass to be 
\begin{equation}
M_{c} \sim 3 \times 10^7 M_{\odot}\,
  \bigg(\frac{T_{\rm{kd}}}{\rm{keV}}\bigg)^{-3/2} \,
  \bigg(\frac{m_{\phi}}{\rm{MeV}}\bigg)^{-3/2}\,.  
\end{equation} 
Combining this expression with our specific particle physics scenario,
we arrive at the estimate: 
\begin{eqnarray}
M_{c} \sim 10^7 M_{\odot} \, \bigg(\frac{m_U}{\rm{MeV}}\bigg)^{-3/2}
\bigg(\frac{m_{\phi}}{\rm{MeV}}\bigg)^{-15/8}  \bigg(\frac{g_{U\phi
    \phi}\, g_{U \nu \nu}}{10^{-6}}\bigg)^{3/4}\,.
\end{eqnarray}

We note that for $T_{\rm kd} \sim {\rm keV}$ and $m_{\phi} \sim {\rm MeV}$, the smallest halos that
 form (those with mass $\sim$$M_c$)
are the ones that host the smallest of the dwarf galaxies seen in the
Milky Way \cite{Strigari:2006rd}. Therefore, the predictions for the 
number of Milky Way satellites will be different in this scenario
compared to that for dark matter with electroweak scale masses. Numerical simulations with
truncated power spectra that are able to resolve halos with masses
below $M_c$  and a detailed treatment of galaxy formation on small
scales will be required to make robust predictions for the satellite
(dwarf) galaxy population in galaxies like the Milky Way and
Andromeda.  

{\bf Relic Abundance and Other Constraints.}
There are a number of constraints on the various
couplings and masses in MeV dark matter. First, we require 
that dark matter is thermally produced in
the early universe with the observed abundance~\cite{wmap}. The annihilation cross section for scalar dark
matter particles through the $s$-channel exchange of a $U$-boson is
\cite{scalar,drees}: 
\begin{eqnarray}
&\sigma v& = \frac{g^2_{U \phi \phi} (s - 4 m^2_{\phi})}{12 \pi s
    [(s-m^2_U)^2 + \Gamma^2_U m^2_U]} \sum_{f} 
\sqrt{1-4 m^2_f/s} \nonumber \\
&\times& [s(g^2_{f_L} + g^2_{f_R})+m^2_f (6 g_{f_L}
    g_{f_R}-(g^2_{f_L} + g^2_{f_R}))] 
\end{eqnarray}
where we have denoted the $U$ couplings
to left and right-handed fermions by $g_{f_L}$ and $g_{f_R}$,
respectively. The sum is over $e^+ e^-$ and the three species of
neutrinos. Notice that at low velocities ($s \approx 4 m^2_{\phi}$)
the cross section approaches zero, being entirely the result of a
P-wave amplitude.  

In order for dark matter annihilations to generate the observed 511
keV photons from the Galactic Bulge, positrons must be injected with
energies no greater than $\sim$ 3 MeV (more energetic positrons
would unacceptably broaden the 511 keV line width). This leads to the
constraint, $0.511 \, \rm{MeV} \lsim m_{\phi} \lsim 3$ MeV
\cite{beacom}. We also require that $m_{\phi} < m_U$ in order to avoid
dark matter annihilating largely to $UU$, which is not s-wave
suppressed~\cite{Jacoby:2007vs}. For a 0.511-3 MeV dark matter
particle to be generated in a quantity consistent with the observed
dark matter abundance, an annihilation cross section on the order of a
picobarn is required during the freeze-out epoch.  

\begin{figure}
\resizebox{8.0cm}{!}{\includegraphics{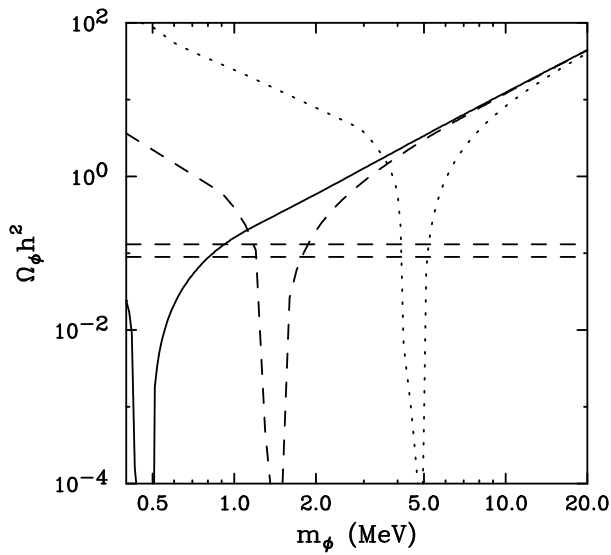}} \\
\caption{The thermal relic abundance of dark matter as a function of
  its mass for $m_U=1$ MeV (solid), 3 MeV (dashed) and 10 MeV
  (dotted). In each case, the product $g_{U\phi \phi} g_{U f f}$ was
  set to $10^{-6}$ for each of $f=e$, $\nu_e$, $\nu_{\mu}$ and
  $\nu_{\tau}$. The dashed horizontal lines denote the measured
  density of dark matter~\cite{wmap}. For illustration, we
  have shown the complete range for the dark matter mass, although only
  $m_{\phi} < m_U$ is viable~\cite{Jacoby:2007vs}.} 
\label{relic}
\end{figure}

We show in Fig.~\ref{relic} the abundance of dark matter in this
model as a function of its mass, for three values of the gauge boson
mass, and for couplings of $g_{U\phi \phi} \times g_{U f f} =
10^{-6}$. Throughout, we adopt a common
$U$-fermion-fermion coupling for electrons and neutrinos. 

The $U$-boson's couplings to fermions are constrained by $\nu e$
scattering experiments such that $g_{U \nu \nu} \sqrt{g^2_{U e_L
    e_L}+g^2_{U e_R e_R}} \lsim m^2_U G_F$~\cite{scalar,fayet}. For
the case of a common $U$-fermion-fermion coupling, this reduces to
$g_{Uff} \lsim  2.9 \times 10^{-6} \times (m_U/\rm{MeV})^2$. A
somewhat weaker constraint can be found from measurements of the
electron's magnetic moment~\cite{scalar}.

\begin{figure}

\resizebox{8.0cm}{!}{\includegraphics{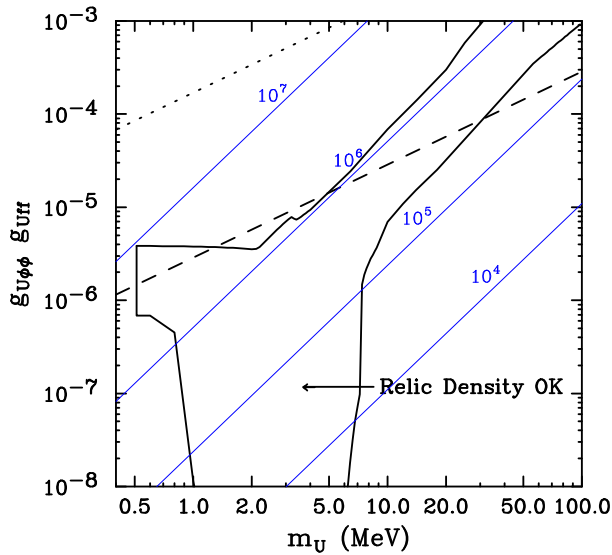}}
\caption{Regions in the $m_U$ versus $g_{U\phi \phi} g_{U f f}$ plane
  in which the measured dark matter density matches the thermal relic
  abundance for some value of $m_{\phi}$ in the range of $m_e$ to 3
  MeV. We have adopted a common $U$-fermion-fermion coupling for
  electrons and neutrinos. The dashed line denotes the constraint from
  $\nu e$ scattering experiments (for the optimal case of $g_{U \phi
    \phi}\approx 1)$~\cite{scalar,fayet}. The dotted line denotes the
  (weaker) constraint from measurements of the electron's magnetic
  moment~\cite{scalar}. The light blue lines are contours of constant
  $M_c$ from $10^4$ to $10^7$ solar masses. Here, we have used
  $m_{\phi}=1$ MeV. From this figure, we see that once all of the
  constraints are considered, $M_c$ in the range of $10^4 \,
  M_{\odot}$ to $10^7$ $M_{\odot}$ are generally expected.} 
\label{plane}
\end{figure}

In Fig.~\ref{plane}, we show the range of $m_U$ and the product
$g_{U\phi \phi} \times g_{U f f}$ for which the measured dark matter
density can be made to match the thermal relic abundance in this model
(for some value of $m_{\phi}$ in the range of $m_e$ to 3 MeV). 
We also show the constraints from $\nu e$
scattering experiments (for the optimal case of $g_{U \phi
  \phi}\approx 1)$ ~\cite{scalar,fayet} and from the measurement of
the electron's magnetic moment \cite{scalar}. 

As light blue lines, we have plotted contours of constant $M_c$ from
$10^7$ to $10^4$ solar masses. Here, we have used $m_{\phi}=1$
MeV. For other masses, the results vary as $M_c \propto
m^{-15/8}_{\phi}$. From this figure, we see that once all of the
constraints are considered, $M_c$ is generally expected to fall in the
range of $10^4$ to $10^7$ $M_{\odot}$. It should be noted that the
region where $m_U \approx 1-6$ MeV and $g_{U\phi \phi} g_{U f f}
\lsim 10^{-7}$ is highly fine tuned and relies on being very close to
the resonance at $2 m_{\phi} \simeq m_U$ to avoid the overproduction
of dark matter. 

The coupling of MeV dark matter to the neutrinos and electrons could
change BBN predictions by causing the neutrino and photon temperatures to be the same to lower redshifts than is standard. Effects of this
nature have been considered previously \cite{bbn},
though it is unclear how these constraints apply to this model on account of the presence of extra thermalized scalars and new neutrino interactions; a detailed analysis of the constraints from BBN is beyond the scope of this letter, but we have checked that there
are viable regions of parameter space where the expected deviations
from standard BBN predictions are within observational bounds
\cite{Nnu}. 

MeV dark matter (and associated $U$-boson) with couplings
to neutrinos would have other observable consequences. The existence
of such a $U$-boson would lead to TeV scale absorption features in the
high-energy cosmic neutrino spectrum~\cite{Hooper:2007jr}. The
spectrum of neutrinos produced in core-collapse supernovae could also be
modified due to their interactions with dark matter particles produced
during the collapse~\cite{sn}. 

{\bf Conclusions.}
In summary, we have calculated the small scale power spectrum of
MeV dark matter. This scenario is motivated by the observation of 511
keV emission from the Galactic Bulge. 
Assuming that the relevant couplings to neutrinos are similar to those to
electrons, we find that MeV dark matter
particles remain in kinetic equilibrium with the cosmic neutrino
background up to temperatures of $\sim$1-10 keV. This late kinetic decoupling leads to larger free-streaming
lengths for MeV dark matter as compared to WIMPs with electroweak
scale masses. This highly suppresses the formation of small
scale structure. Depending on the parameters considered,
the matter power spectrum is expected to be cutoff below $10^7\,
M_{\odot}$ to $10^4\, M_{\odot}$ in this scenario.  

This result has a number of particularly interesting astrophysical
implications. First, it predicts a cutoff in the mass function of
dwarf galaxies at a mass scale much larger than that for WIMPs with electroweak scale masses. 
It was previously shown that the number of
dwarf galaxy-sized dark matter halos in numerical simulations of
cold dark matter is considerably larger than the observed populations
in the Milky Way and Andromeda galaxies ({\it ie.} the ``missing
satellites problem")~\cite{Klypin:1999uc,Moore:1999nt}. 
This issue may be resolved by astrophysical means~\cite{astrosolution}  
or by altering the nature of the dark matter's interactions~\cite{int}
or mechanism of production~\cite{pro}.
The model we consider falls into this latter category, though detailed
numerical simulations will be required to make precise predictions of
the satellite population in  MeV dark matter. 

Tests of dark matter models with observations of small scale
structure, as we have discussed in this paper, are becoming a more
realistic possibility given the recent discoveries of faint satellite
companions to the Milky Way and Andromeda galaxies~\cite{new}. Present
estimates of the masses of these new satellites fall in the range
$10^5 \, M_{\odot}$ to $10^7 \, M_{\odot}$, which is near the cutoff
mass scale in MeV dark matter. Present and future searches for faint
satellites, and the characterization of the mass function at these
scales, will thus provide important constraints on MeV dark matter.

\smallskip
We thank John Beacom for discussions on this topic. 
This work has been supported by the US Department of Energy, including
grant DE-FG02-95ER40896, and by NASA grant NAG5-10842, and by NSF
grants AST-0607746 and PHY-0555689. We acknowledge the Aspen
center for Physics where this work was initiated.


\begin{thebibliography}{}




\bibitem{wimp}
  X.~l.~Chen, M.~Kamionkowski and X.~Zhang,
  Phys.\ Rev.\  D {\bf 64}, 021302 (2001).
  V.~Berezinsky, V.~Dokuchaev and Y.~Eroshenko,
  Phys.\ Rev.\  D {\bf 68}, 103003 (2003);
  A.~M.~Green, S.~Hofmann and D.~J.~Schwarz,
  Mon.\ Not.\ Roy.\ Astron.\ Soc.\  {\bf 353}, L23 (2004);
  JCAP {\bf 0508}, 003 (2005);
  S.~Profumo, K.~Sigurdson and M.~Kamionkowski,
  Phys.\ Rev.\ Lett.\  {\bf 97}, 031301 (2006);


\bibitem{loeb}
  A.~Loeb and M.~Zaldarriaga,
  Phys.\ Rev.\  D {\bf 71}, 103520 (2005).


\bibitem{mangano}
  G.~Mangano, {\it et al.},
  Phys.\ Rev.\  D {\bf 74}, 043517 (2006);
  C.~Boehm, {\it et al.},
  arXiv:astro-ph/0309652.

\bibitem{integral}
  P.~Jean {\it et al.},
  Astron.\ Astrophys.\  {\bf 407}, L55 (2003).

\bibitem{511dark}
 C.~Boehm, D.~Hooper, J.~Silk, M.~Casse and J.~Paul,
  Phys.\ Rev.\ Lett.\  {\bf 92}, 101301 (2004).



\bibitem{scalar}
  C.~Boehm and P.~Fayet,
  Nucl.\ Phys.\ B {\bf 683}, 219 (2004).


\bibitem{fayet}
  P.~Fayet,
  Phys.\ Rev.\ D {\bf 70}, 023514 (2004).

\bibitem{drees}
  N.~Borodatchenkova, D.~Choudhury and M.~Drees,
  Phys.\ Rev.\ Lett.\  {\bf 96}, 141802 (2006).

\bibitem{collider}
  B.~McElrath,
  Phys.\ Rev.\ D {\bf 72}, 103508 (2005);
  P.~Fayet,
  arXiv:hep-ph/0607094;
  M.~Ablikim {\it et al.}  [BES Collaboration],
  Phys.\ Rev.\ Lett.\  {\bf 97}, 202002 (2006);
  P.~Fayet,
  Phys.\ Rev.\ D {\bf 74}, 054034 (2006).


\bibitem{boehmneutrino}
  C.~Boehm,
  Phys.\ Rev.\ D {\bf 70}, 055007 (2004).



\bibitem{atomic}
  C.~Bouchiat and P.~Fayet,
  Phys.\ Lett.\ B {\bf 608}, 87 (2005).


\bibitem{sn}
  P.~Fayet, D.~Hooper and G.~Sigl,
  Phys.\ Rev.\ Lett.\  {\bf 96}, 211302 (2006).



\bibitem{beacom}
  J.~F.~Beacom and H.~Yuksel,
  Phys.\ Rev.\ Lett.\  {\bf 97}, 071102 (2006);
  J.~F.~Beacom, N.~F.~Bell and G.~Bertone,
  Phys.\ Rev.\ Lett.\  {\bf 94}, 171301 (2005).


\bibitem{bbn}
  E.~W.~Kolb, M.~S.~Turner and T.~P.~Walker,
  Phys.\ Rev.\  D {\bf 34}, 2197 (1986);
 P.~D.~Serpico and G.~G.~Raffelt,
  Phys.\ Rev.\  D {\bf 70}, 043526 (2004); 

\bibitem{wmap}
  D.~N.~Spergel {\it et al.}  [WMAP Collaboration],
  arXiv:astro-ph/0603449.

\bibitem{boltz}
  E.~Bertschinger,
  Phys.\ Rev.\  D {\bf 74}, 063509 (2006);
  T.~Bringmann and S.~Hofmann,
  arXiv:hep-ph/0612238.


\bibitem{lymanalpha}
  U.~Seljak, {\it et al.},
  Phys.\ Rev.\ Lett.\  {\bf 97}, 191303 (2006);
  M.~Viel, {\it et al.},
  Phys.\ Rev.\ Lett.\  {\bf 97}, 071301 (2006).


\bibitem{Strigari:2006rd}
  L.~E.~Strigari, S.~M.~Koushiappas, J.~S.~Bullock and M.~Kaplinghat,
  arXiv:astro-ph/0611925.

\bibitem{Jacoby:2007vs}
  C.~Jacoby and S.~Nussinov,
  arXiv:hep-ph/0703014.
  
\bibitem{Nnu}
  R.~H.~Cyburt, B.~D.~Fields, K.~A.~Olive and E.~Skillman,
  Astropart.\ Phys.\  {\bf 23}, 313 (2005)
  [arXiv:astro-ph/0408033].

  \bibitem{Hooper:2007jr}
  D.~Hooper,
  arXiv:hep-ph/0701194;
 S.~Palomares-Ruiz and T.~Weiler, in progress.


\bibitem{Klypin:1999uc}
  A.~A.~Klypin, {\it et al.},
  Astrophys.\ J.\  {\bf 522}, 82 (1999).

\bibitem{Moore:1999nt}
  B.~Moore, {\it et al.},
  Astrophys.\ J.\  {\bf 524}, L19 (1999).

\bibitem{astrosolution}
  J.~S.~Bullock, A.~V.~Kravtsov and D.~H.~Weinberg,
  Astrophys.\ J.\  {\bf 539}, 517 (2000);
  A.~V.~Kravtsov, O.~Y.~Gnedin and A.~A.~Klypin,
  Astrophys.\ J.\  {\bf 609}, 482 (2004);
  B.~Moore, {\it et al.},
  Mon.\ Not.\ Roy.\ Astron.\ Soc.\  {\bf 368}, 563 (2006).

\bibitem{int}
  C.~Boehm, P.~Fayet and R.~Schaeffer,
  Phys.\ Lett.\  B {\bf 518}, 8 (2001);
  X.~l.~Chen, S.~Hannestad and R.~J.~Scherrer,
  Phys.\ Rev.\  D {\bf 65}, 123515 (2002).



\bibitem{pro}
  K.~Sigurdson and M.~Kamionkowski,
  Phys.\ Rev.\ Lett.\  {\bf 92}, 171302 (2004);
  J.~A.~R.~Cembranos, {\it et al.},
  Phys.\ Rev.\ Lett.\  {\bf 95}, 181301 (2005);
  M.~Kaplinghat,
  Phys.\ Rev.\  D {\bf 72}, 063510 (2005).

\bibitem{new}
B.~Willman {\it et al.},
Astrophys.\ J.\ {\bf 626}, L85 (2005);
D.~B.~Zucker {\it et al.}Ê [SDSS Collaboration],
Astrophys.\ J.\ {\bf 643}, L103 (2006);
V.~Belokurov {\it et al.}Ê [SDSS Collaboration],
Astrophys.\ J.\ {\bf 654}, 897 (2007).



\end{thebibliography}
\end{document}